\documentclass[12pt]{article}
\usepackage{amsmath}
\usepackage{amssymb}

\newcommand{\BE}{\begin{equation}}
\newcommand{\EE}{\end{equation}}

\begin{document}

\setcounter{page}{0} \thispagestyle{empty}

%\begin{flushright}
%{\small BARI-TH .../2006}
%\end{flushright}

\vspace*{0.5cm}

\begin{center}
{\large \bf On the low-energy spectrum of spontaneously broken
$\Phi^4$ theories }

\end{center}

\vspace*{1.5cm}

\renewcommand{\thefootnote}{\fnsymbol{footnote}}

\begin{center}
{\large M. Consoli}
 \\[0.3cm]
 {INFN - Sezione di Catania, I-95123 Catania, Italy}
\end{center}

\vspace*{0.5cm}

\vspace*{1.0cm}

\renewcommand{\abstractname}{\normalsize Abstract}
\begin{abstract}
The low-energy spectrum of a one-component, spontaneously broken
$\Phi^4$ theory is generally believed to have the same simple
massive form $\sqrt{{\bf p}^2 + m^2_h}$ as in the symmetric phase
where $\langle \Phi \rangle=0$. However, in lattice simulations of
the 4D Ising limit of the theory, the two-point connected correlator
and the connected scalar propagator show deviations from a standard
massive behaviour that do not exist in the symmetric phase. As a
support for this observed discrepancy, I present a variational,
analytic calculation of the energy spectrum $E_1({\bf p})$ in the
broken phase. This analytic result, while providing the trend
$E_1({\bf p})\sim \sqrt{{\bf p}^2 + m^2_h}$ at large $|{\bf p}|$,
gives an energy gap $E_1(0)< m_h$, even when approaching the
infinite-cutoff limit $\Lambda \to \infty$ with that infinitesimal
coupling $ \lambda \sim 1/\ln \Lambda$ suggested by the standard
interpretation of ``triviality" within leading-order perturbation
theory. I also compare with other approaches and discuss the more
general implications of the result.
%\vskip 10 pt PACS number(s): 11.30.Qc, 14.80.Bn

\end{abstract}

\vspace*{0.5cm}
\begin{flushleft}
%PACS number(s): 11.30.Qc, 14.80.Bn, 11.15.Ha.
\end{flushleft}
\renewcommand{\thesection}{\normalsize{\arabic{section}.}}
\vfill\eject
%%%%%%%%%%%%%%%%%%%%%%%%%%%%%%%%%%%%%%%%%%%%%%%%%%%%%%%%%%%%%%%%%%%%%%%
\section{\normalsize{Introduction}}
%\label{Introduction}
%%%%%%%%%%%%%%%%%%%%%%%%%%%%%%%%%%%%%%%%%%%%%%%%%%%%%%%%%%%%%%%%%%%%%%%

In the case of a one-component, spontaneously broken $\Phi^4$
theory, one usually assumes a form of single-particle energy
spectrum, say $E_1({\bf p})=\sqrt{{\bf p}^2 + m^2_h}$, as in a
simple massive theory with no qualitative difference from the
symmetric phase where $\langle \Phi \rangle=0$.

One can objectively test \cite{further} this expectation with
lattice simulations, performed in the 4D Ising limit of the theory,
and study the exponential decay of the connected two-point
correlator $C_1({\bf p}, t)\sim e^{-E_1({\bf{p}})t}$ and the
connected scalar propagator $G(p)$. Differently from the symmetric
phase, where the simple massive picture works to very high accuracy,
the results of the low-temperature phase show unexpected deviations.
Namely, when the 3-momentum ${\bf p} \to 0$, the fitted
$E_1({\bf{p}})$ deviates from (the lattice version of) the standard
massive form $\sqrt { {\bf p}^2 + {\rm const.} }$~ and, when the
4-momentum $p_\mu\equiv ({\bf p},p_4) \to 0$, the measured $G(p)$
deviates from (the lattice version of) the form $1/(p^2+ {\rm
const.})$ .

After the first indications of Ref.\cite{further}, Stevenson
\cite{stevensonnp} checked independently the existence of this
discrepancy in the lattice data of other authors. To this end, he
started from the lattice data of Ref.\cite{balog} for the time
slices of $C_1({\bf p}=0, t)$ and used the Fourier-transform
relation to generate equivalent data for the connected scalar
propagator $G(p)$. The resulting behaviour of $G(p)$ is in complete
agreement with the analogous plots obtained from Ref.\cite{further}
(compare Figs.6c, 7, 8 and 9 of Ref.\cite{stevensonnp}).

The whole issue was later re-considered in Ref.\cite{step}.
According to these authors, at the present, after taking into
account various theoretical uncertainties, the deviations are not so
statistically compelling. In their opinion, the conventional
scenario of a simple, weakly coupled, massive theory, "unfortunately
can only be nailed down by analytic proofs".

The aim of this Letter is to present, in Sects.2 and 3, a possible
analytic proof under the form of a variational calculation of the
energy spectrum in the broken-symmetry phase. This analytic result,
while indeed providing a behaviour $E_1({\bf p})\sim \sqrt{{\bf p}^2
+ m^2_h}$ at larger $|{\bf p}|$, gives theoretical support for
deviations in the ${\bf p}\to 0$ limit. In particular, the energy
gap $E_1(0)$ is definitely smaller than the $m_h$ parameter that
enters the asymptotic form of the spectrum. I emphasize that the
estimate, being of variational nature, constrains from above the
ratio ${{E_1(0)}\over{m_h}}$ whose value, by enlarging the
variational subspace, can only decrease. In addition, the result
persists when taking the infinite cutoff limit $\Lambda \to \infty$
with the typical trend of the coupling constant $\lambda \sim 1/\ln
\Lambda$ that is expected in the standard interpretation of
``triviality'' \cite{triviality} within leading-order perturbation
theory.  Finally, in Sect.4, I will also compare with other
approaches and discuss the more general implications of the result.\

\section{\normalsize{Stability analysis of $\Phi^4$ theory}}

The preliminary starting point, for any variational calculation in
the broken-symmetry phase of a one-component $\Phi^4$ theory, is the
basic Hamiltonian operator ($\lambda >0$) \BE \label{normal}
\hat{H}= ~:\int d^3x\left[ {{1}\over{2}} \left( \Pi^2 +
(\nabla\Phi)^2 + \Omega^2_o \Phi^2 \right) +
{{\lambda}\over{4!}}\Phi^4\right]:\EE where ( $\omega_{\bf
k}(\Omega)=\sqrt{{\bf k}^2 + \Omega^2}$)\BE \Phi({\bf x})= \int
{{d^3 k }\over{\sqrt {2 \omega_{\bf k}(\Omega_o) (2\pi)^3} }} \left(
a_{\bf k} \exp  i {\bf k}\cdot{\bf x} + a^{\dagger}_{\bf k} \exp -i
{\bf k}\cdot{\bf x} \right)\EE and \BE \Pi({\bf x})= i\int {{d^3
k}\over{ \sqrt{(2\pi)^3} }}  \sqrt{ {{ \omega_{\bf k}(\Omega_o) }
\over{2 }} } \left( a^{\dagger}_{\bf k} \exp -i {\bf k}\cdot{\bf x}
- a_{\bf k} \exp i {\bf k}\cdot{\bf x} \right)\EE In
Eq.(\ref{normal}) normal ordering is defined with respect to a
reference state $|0\rangle$ which is the vacuum of the creation and
annihilation operators ($ a_{\bf k}|0\rangle =\langle
0|a^{\dagger}_{\bf k}=0$) with commutation relations $ [a_{\bf
k},a^{\dagger}_{\bf k'}]=\delta^{(3)}({{\bf k}-{\bf k'}})$. The
standard stability analysis for the above Hamiltonian is performed
in the class of the normalized gaussian ground states
$|\Psi^{(0)}\rangle \equiv |\Psi^{(0)}(\Omega,\varphi)\rangle$  with
\cite{gaussian1,gaussian2} \BE \label{phi1} \langle\Psi^{(0)} |\Phi|
\Psi^{(0)}\rangle = \varphi \EE and \BE\label{phi2} \langle
\Psi^{(0)}|\Phi({\bf x}) \Phi({\bf y})| \Psi^{(0)}\rangle=\varphi^2
+ G({\bf x},{\bf y}) \EE where \BE \label{phi3} G({\bf x},{\bf y})=
\int {{d^3 k }\over{2 \omega_{\bf k}(\Omega) (2\pi)^3}} \exp i {\bf
k}\cdot({\bf x}-{\bf y}) \EE is the equal-time propagator of the
shifted fluctuation field \BE h({\bf x}) = \Phi({\bf x})-\varphi \EE
with \BE h({\bf x})= \int {{d^3 k }\over{\sqrt {2 \omega_{\bf
k}(\Omega) (2\pi)^3} }} \left( b_{\bf k} \exp  i {\bf k}\cdot{\bf x}
+ b^{\dagger}_{\bf k} \exp -i {\bf k}\cdot{\bf x} \right) \EE Thus,
the relation with the reference vacuum state is $|0\rangle\equiv
|\Psi^{(0)}(\Omega_o,\varphi=0)\rangle$ at which $b_{\bf k}\equiv
a_{\bf k}$. Equivalently, one could switch to a functional formalism
where the gaussian ground states are described by the class of
functionals \cite{ciancitto} \BE \Psi^{(0)}[\Phi]=({\rm Det}~
G)^{-1/4}\exp {-{{1}\over{4}}\int d^3x\int d^3y (\Phi({\bf
x})-\varphi)G^{-1}({\bf x},{\bf y})(\Phi({\bf y})-\varphi)} \EE In
this equivalent approach, the field operator $\Phi({\bf x})$ acts on
$\Psi^{(0)}[\Phi]$ multiplicatively while the momentum operator acts
by functional differentiation \BE \Pi({\bf x})\Psi^{(0)}[\Phi]
={{1}\over{i}} {{\delta}\over{\delta \Phi({\bf x})
}}\Psi^{(0)}[\Phi] \EE In the following, I shall maintain the
standard second-quantized representation (1)-(8) for its more
intuitive character.

As shown in Ref.\cite{ciancitto}, the states
$|\Psi^{(0)}(\Omega,\varphi)\rangle$ can be represented as coherent
states built up with the original $a_{\bf k}$ and $a^{\dagger}_{\bf
k}$ operators. In this sense, they represent forms of condensed
vacua and the old operators are related to the new "quasiparticle"
 $b_{\bf k}$ and $b^{\dagger}_{\bf k}$ operators (whose vacuum is
 $|\Psi^{(0)}(\Omega,\varphi)\rangle$) by a Bogolubov
transformation that includes a shift of the zero-momentum mode.

The expectation value of the Hamiltonian in the class of the
gaussian ground states gives the gaussian energy density
$W_G(\varphi,\Omega)$ \BE \langle\Psi^{(0)} |\hat{H}|
\Psi^{(0)}\rangle=\int d^3x~ (W_G(\varphi,\Omega)-
W_G(0,\Omega_o))\EE where ($I_o(\Omega)=G({\bf x},{\bf x})$,
$I_1(\Omega) = {{1}\over{8}}G^{-1}({\bf x},{\bf x})$~) \BE
W_G(\varphi,\Omega)= I_1(\Omega) + {{1}\over{2}} m^2_B \varphi^2 +
{{\lambda}\over{4!}}\varphi^4 +{{1}\over{2}}\left( m^2_B +
{{\lambda}\over{2}}\varphi^2 -\Omega^2 +
{{\lambda}\over{4}}I_o(\Omega)\right)I_o(\Omega) \EE and, just for
simplicity of notation, the quantity \BE \label{bare} m^2_B\equiv
\Omega^2_o-{{\lambda}\over{2}}I_o(\Omega_o)\EE has been introduced.
It plays the role of a `bare mass' for the quantum theory but his
origin depends on the normal ordering prescription adopted for the
Hamiltonian Eq.(\ref{normal}).

Now, the existence of the $\Phi^4$ critical point \cite{glimm}
implies that, for sufficiently large and negative values of $m^2_B$,
the cutoff theory will exhibit spontaneous symmetry breaking. In
this regime, one can explore the conditions for non-trivial minima
with $\varphi \neq 0$. Minimization of $W_G$ with respect to
$\varphi$ gives \BE \label{phi_B} {{\partial
W_G(\varphi,\Omega)}\over{\partial \varphi}}= \varphi\left( m^2_B +
{{\lambda}\over{6}}\varphi^2 +
{{\lambda}\over{2}}I_o(\Omega)\right)=0\EE while minimization with
respect to $\Omega$ yields \BE \label{omegap}
\Omega^2(\varphi)=m^2_B + {{\lambda}\over{2}}\varphi^2 +
{{\lambda}\over{2}}I_o(\Omega) \EE Finally, the replacement $\Omega=
\Omega(\varphi)$ in $W_G(\varphi,\Omega)$ provides the gaussian
effective potential (GEP) \BE
V_G(\varphi)=W_G(\varphi,\Omega(\varphi))-W_G(0,\Omega_o)\EE By
combining Eqs.(\ref{phi_B}) and (\ref{omegap}), non-trivial extrema
$\varphi\neq 0$ can only occur at those values $\varphi=\pm v$ where
\BE \label{omegav} \Omega^2(v)={{\lambda}\over{3}}v^2\equiv m^2_h
\EE The standard identification of $m_h$ with the energy-gap of the
broken phase derives from the following argument. At the absolute
minima $\varphi=\pm v$, the same Hamiltonian in Eq.(\ref{normal})
becomes also normal ordered in the creation and annihilation
operators $b_{\bf p}$ and $b^{\dagger}_{\bf p}$ \cite{ciancitto},
namely one finds \BE \label{normal1} \hat{H}= E_0+ \hat{H}_2 +
\hat{H}_{\rm int} \EE Here \BE \label{ground} E_o= \int d^3x~ V_G(v)
< 0\EE is the gaussian ground-state energy. The quadratic operator
\BE \hat{H}_2= \int d^3p~\omega_{\bf p}(m_h)~b^{\dagger}_{\bf
p}b_{\bf p} \EE describes free-field quanta with energies
$\omega_{\bf p}(m_h)=\sqrt{{\bf p}^2 + m^2_h}$~ and finally \BE
\hat{H}_{\rm int}= \int d^3x :\left({{\lambda v}\over{3!}} h^3({\bf
x}) + {{\lambda}\over{4!}} h^4({\bf x})\right): \EE takes into
account the residual self-interactions that have not been reabsorbed
into the vacuum structure and in the mass parameter $m_h$. In the
above relation, normal ordering of the $b^{\dagger}_{\bf p}$ and
$b_{\bf p}$ operators is now defined with respect to one of the two
equivalent absolute minima of the GEP for $\varphi=\pm v$. In this
way, by introducing the one-quasiparticle states (see Eq.(6.4) of
Ref.\cite{gaussian2}) \BE |1,{\bf p} \rangle =b^{\dagger}_{\bf
p}|\Psi^{(0)}\rangle \sqrt {2 \omega_{\bf p} (2\pi)^3} \EE one finds
\BE {{  \langle 1,{\bf p}|(\hat{H}-E_0)|1,{\bf p}
\rangle}\over{\langle 1,{\bf p}|1,{\bf p} \rangle }}=\sqrt{{\bf p}^2
+ m^2_h} \EE and it becomes natural to identify $m_h$ with the
energy-gap of the broken phase. In the following section, I will
check this expectation with a variational calculation.

\section{\normalsize{Variational calculation of the energy gap in the broken phase}}

The variational procedure is of the same type considered by Di Leo
and Darewych \cite{dileo} and by Siringo \cite{siringo} when
discussing the bound-state problem in the Higgs sector, namely \BE
|\Psi_1 \rangle= A({\bf{q}})b^{\dagger}_{\bf q}|\Psi^{(0)}\rangle +
\int
d^3k~B({\bf{k}},{\bf{q}})b^{\dagger}_{\bf{k+q}}b^{\dagger}_{{\bf
-k}}|\Psi^{(0)}\rangle \EE with $B({\bf{k}},{\bf{q}})= B({- \bf{k}}-
{\bf{q}},{\bf{q}})$.

The two complex functions $A({\bf{q}})$ and $B({\bf{p}},{\bf{q}})$
have to be determined in order to solve the eigenvalue problem for
the Hamiltonian $\hat{H}$ Eq.(\ref{normal1}) in the chosen subspace.
By denoting with ${E}_1={E}_1({\bf{q}})$ the corresponding
eigenvalue, one gets coupled equations (everywhere $\omega_{\bf
p}=\omega_{\bf p}(m_h)$) \BE \label{Aq} {{\delta\langle
\Psi_1|(\hat{H}-E_o-{E}_1)|\Psi_1 \rangle }\over{\delta
A^*({\bf{q}}) }}=A({\bf{q}})(\omega_{\bf q}-{E}_1) + f({\bf{q}})=0
\EE and \BE\label{Bkq} {{\delta\langle
\Psi_1|(\hat{H}-E_o-{E}_1)|\Psi_1 \rangle }\over{\delta
B^*({\bf{k}},{\bf{q}}) }}= 2B({\bf{k}},{\bf{q}})[\omega_{\bf k}+
\omega_{\bf k+q }-{E}_1] + g({\bf{k}},{\bf{q}})=0 \EE In
Eqs.(\ref{Aq}) and (\ref{Bkq}) $f({\bf{q}})$ and
$g({\bf{k}},{\bf{q}})$ are defined as \BE f({\bf{q}})= {{\lambda
v}\over{ 8 \pi^{3/2} \sqrt{\omega_{\bf q} }}} \int d^3k~{{
B({\bf{k}},{\bf{q}}) }\over{ \sqrt { \omega_{\bf k} \omega_{\bf k+q}
}  }} \EE and \BE g({\bf{k}},{\bf{q}})={{\lambda v}\over{ 8
\pi^{3/2} \sqrt{\omega_{\bf q} } }} {{A({\bf{q}})}\over{
\sqrt{\omega_{\bf k}\omega_{\bf k+ q} } }} + {{\lambda }\over{ 32
\pi^{3} \sqrt{ \omega_{\bf k}\omega_{\bf {k+ q}} } }}  \int d^3p~{{
B({\bf{p}},{\bf{q}}) }\over{ \sqrt { \omega_{\bf p}\omega_{\bf
{p+q}}} }} \EE The two functions $f({\bf{q}})$ and
$g({\bf{k}},{\bf{q}})$ contain the same integral up to numerical
factors. This allows to eliminate {\it exactly}
$B({\bf{k}},{\bf{q}})$ in favour of $A({\bf{q}})$  as \BE
B({\bf{k}},{\bf{q}})= {{A({\bf{q}})}\over{ 8 v~ \pi^{3/2} }} \sqrt{
{{\omega_{\bf q} }\over{\omega_{\bf k} \omega_{\bf k+q} }} }  \left(
{{ \omega_{\bf q}-E_1 - {{3m^2_h}\over{2\omega_{\bf q}}}
}\over{\omega_{\bf k}+ \omega_{\bf k+q }-E_1}} \right) \EE after
using the relation (\ref{omegav}) $m^2_h={{\lambda v^2}\over{3}}$.
By replacing in Eq.(\ref{Aq}), one obtains \BE A({\bf{q}})
(\omega_{\bf q}-E_1) + A({\bf{q}}) \left(\omega_{\bf q}-E_1 -
{{3m^2_h}\over{2\omega_{\bf q}}}\right) ~{{\lambda}\over{16 \pi^2}}
J({\bf{q}})=0 \EE where \BE \label {Jk} J({\bf{q}})=
{{1}\over{4\pi}} \int {{d^3p}\over{\omega_{\bf p}\omega_{\bf p+
q}[\omega_{\bf p}+ \omega_{\bf p+ q} -E_1({\bf{q}})] }} \EE
Therefore, for $A({\bf{q}}) \neq 0$, one obtains the final relation
for the eigenvalue \BE\label{eigenfinal} {E}_1({\bf{q}})=\omega_{\bf
q}\left(1 - {{3m^2_h}\over{2\omega^2_{\bf q}}} F({\bf{q}}) \right)
\EE where \BE F({\bf{q}})=  {{ {{\lambda}\over{16 \pi^2}}
J({\bf{q}}) }\over{ 1 +{{\lambda}\over{16 \pi^2}} J({\bf{q}}) }} \EE
Now, the integral in Eq.(\ref{Jk}) diverges logarithmically  \BE J
\sim \int^{\Lambda}_0 {{ p^2 dp} \over{ 2(p^2 + m^2_h)^{3/2} }} \sim
{{1}\over{2}}\ln {{\Lambda}\over{m_h}} \EE so that any conclusion on
the energy spectrum depends on the possible behaviours of the
coupling constant $\lambda$ when the ultraviolet cutoff $\Lambda \to
\infty$. A straightforward $\Lambda \to \infty$ limit for
$\lambda=$fixed would yield $F({\bf{q}}) \to 1$ and a negative
$E_1(0)$. However, a more meaningful continuum limit could be
obtained, for instance, by interpreting $\lambda$ as the value of a
running coupling $\lambda(\mu)$ at some scale $\mu$ and then
requiring $\lambda(\mu) \sim 1/\ln(\Lambda/\mu)$ as suggested by the
standard interpretation of ``triviality" within leading-order
perturbation theory.

For a self-consistent derivation of this trend within our
Hamiltonian formalism, let us return to equation (\ref{omegap}) and
use relation (\ref{bare}) to replace the bare mass. For simplicity,
I shall first consider the case $\Omega_o=0$, i.e. \BE m^2_B= -
{{\lambda}\over{2}}I_o(0) \EE By using the identity of
Ref.\cite{gaussian2} \BE I_o(\Omega) - I_o(0)=
-{{\Omega^2}\over{8\pi^2}} \left( \ln {{\Lambda}\over{\Omega}}
+{{1}\over{2}}\right) \EE equation (\ref{omegap}) for $\varphi=\pm
v$, where $\Omega$ is given in Eq.(\ref{omegav}), reduces to the
relation \BE \label{choice} 1= {{\lambda}\over{8\pi^2}}\left( \ln
{{\Lambda}\over{m_h}} +{{1}\over{2}} \right) \EE One can give
different interpretations to this equation. On the one hand, if
$\Phi^4$ theory were just considered a cutoff theory, it might
simply express $m_h$ in terms of the two basic, fixed parameters
$\lambda$ and $\Lambda$. On the other hand, in a Renormalization
Group (RG) perspective, it could also be used to determine a
suitable flow of the coupling constant $\lambda= \lambda(\Lambda)$,
in the two-parameter $(\lambda,\Lambda)$ space, that corresponds to
the same value of $m_h$. As anticipated, from this latter RG point
of view and within leading-order perturbation theory, the resulting
trend $\lambda\sim 1/\ln{{\Lambda}\over{m_h}}$ would be similar to
the $ \Lambda-$dependence of the "renormalized" coupling
$\lambda_R$, usually identified with the value of a running coupling
$\lambda(\mu)$ at a typical finite scale $\mu \sim m_h$. However, in
principle, $\lambda$ might also be considered a "bare" coupling
$\lambda_B$, and thus identified with a running coupling
$\lambda(\mu)$ at an asymptotic ultraviolet scale $\mu\sim \Lambda$.
 As discussed in Ref.\cite{trivpert}, this latter point of
view cannot be ruled out. In fact, the trend $\lambda_B\sim 1/\ln
\Lambda$ represents a completely consistent solution that yields
"triviality" (i.e. $\lambda_R=0$) to any finite order in
perturbation theory by avoiding the problems posed by the 1-loop,
3-loop, 5-loop,.. Landau poles and by the 2-loop, 4-loop,...
spurious ultraviolet fixed points at finite coupling that arise in
the conventional interpretation. In the more general context of the
$\epsilon-$expansion, these two distinct points of view might also
reflect the existence of two separate $\Phi^4$ theories inhabiting
in $d=4+\epsilon$ and $d=4-\epsilon$ space-time dimensions
\cite{zeit}.

In any case, regardless of these interpretative aspects, the
consistency of the whole calculation requires to adopt
Eq.(\ref{choice}) to fix the $(\lambda,\Lambda,m_h)$
interdependence. In this way, one can control the ultraviolet
divergence in $J({\bf{q}})$ and obtain a finite value for
$F({\bf{q}})$. Notice however that, independently of the given
finite value of $F({\bf{q}})$, one gets \BE \label{asye 1}
E_1({\bf{q}})\sim \sqrt{ {\bf q}^2 + m^2_h} \EE at large $|{\bf q}|$
and \BE E_1(0) < m_h \EE consistently with $J(0)$ and $F(0)$ being
positive-definite quantities for any $E_1(0) < 2 m_h$.

The numerical estimate of the energy gap can be obtained from the
relation \BE E_1(0)= m_h\left(1 - {{3}\over{2}}~
 {{ {{\lambda}\over{16 \pi^2}}
J(0) }\over{ 1 +{{\lambda}\over{16 \pi^2}} J(0) }}\right) \EE with
\BE J(0)= \int^{\Lambda}_0 {{p^2 dp}\over{ (p^2 + m^2_h) [ 2
\sqrt{p^2 + m^2_h} - E_1(0)] }} \EE Thus, by defining $p= m_h \sinh
t$ and introducing $\epsilon_1\equiv E_1(0)/m_h$, one obtains \BE
J(0)= \int^{t_{\rm max}}_0 {{\sinh^2t~ dt}\over{ \cosh t[ 2 \cosh t
-\epsilon_1 ] }} \EE or \BE J(0)= {{t_{\rm max} }\over{2}} +
{{1}\over{\epsilon_1}} \left[ {{\pi}\over{2}} - \sqrt{ 1-
{{\epsilon^2_1}\over{4}} } \left(\arcsin(\epsilon_1/ 2)
+{{\pi}\over{2}} \right) \right] \EE where $t_{\rm max}= \ln (2
\Lambda/m_h)$. In this way, in a double limit $t_{\rm max} \to
\infty$ and $\lambda \to 0$, such that $\lambda t_{\rm max}$ is
finite, $\epsilon_1$ is definitely smaller than unity. With the
trend in Eq.(\ref{choice}), one finds ${{\lambda}\over{16 \pi^2}}
J(0)= 1/4 +{\cal O}({{1}\over{t_{\rm max}}} )$ or $F(0)=1/5 +{\cal
O}({{1}\over{t_{\rm max}}} )$ so that \BE \label{varia}
{{E_1(0)}\over{m_h}}=\epsilon_1=0.7 \left(1 + {\cal
O}({{1}\over{t_{\rm max}}})\right) \EE In the same approximation,
where also $F({\bf{q}})-F(0)$ represents a non-leading ${\cal
O}({{1}\over{t_{\rm max}}} )$ effect, the form of the spectrum
becomes very simple and one finds \BE
\label{leading}E_1({\bf{q}})\sim \omega_{\bf q} - {{3}\over{10}}
{{m^2_h}\over{\omega_{\bf q}}}  \EE I emphasize that the result in
Eq.(\ref{varia}) is of variational nature. Therefore, by maintaining
the same relation  Eq.(\ref{choice}) for the coupling constant, and
by enlarging the variational subspace for the Hamiltonian
Eq.(\ref{normal1}) to include higher-order components
$|b^{\dagger}b^{\dagger}b^{\dagger}\rangle$,
$|b^{\dagger}b^{\dagger}b^{\dagger}b^{\dagger}\rangle$,..., the
ratio ${{E_1(0)}\over{m_h}}$ can only decrease.

Exactly the same procedure can be repeated in the more general case
where the $\Omega_o$ mass parameter of the symmetric phase  is non
vanishing. As one can check, by requiring the broken phase to
represent anyway the absolute minimum of the gaussian effective
potential, Eq.(\ref{choice}) can only be modified up to non-leading
${\cal O}(\lambda)$ terms. As a consequence, Eq.(\ref{varia}) is
also modified up to non-leading ${\cal O}({{1}\over{t_{\rm max}}})$
terms and the basic result remains unaffected.

Before  concluding this section, I have to explain the considerable
differences between the conclusions of the present Letter and those
of Ref.\cite{siringo}. There, the analysis was performed directly in
the broken-symmetry phase without considering the overall stability
of the basic $\Phi^4$ Hamiltonian (\ref{normal}) in the class of the
gaussian ground states. For this reason, there was no obvious
guiding principle to relate $\lambda$ to the ultraviolet cutoff
$\Lambda$ and to $m_h$ as in Eq.(\ref{choice}). Thus, differently
from the approach followed in the present Letter, one could try to
take the $\Lambda \to \infty$ limit at $\lambda=$ fixed in such a
way that $\lambda J(0) \to \infty$ and \BE F(0)=  {{
{{\lambda}\over{16 \pi^2}} J(0) }\over{ 1 +{{\lambda}\over{16
\pi^2}} J(0) }} \to 1 \EE In this framework, it was adopted a
particular mass renormalization condition (see Eqs.(20), (21), (27)
and (30) of Ref.\cite{siringo}) \BE \delta m^2=- \lambda v^2 F(0)
\EE in order to get, in the broken-symmetry phase, an exactly free
massive spectrum up to terms that vanish in the $\Lambda\to \infty$
limit. Now, it would be very hard to understand the choice of such a
vacuum-dependent mass counterterm in the context of the basic
Hamiltonian (\ref{normal}). Therefore, it should not come as a
surprise that, by changing the renormalization conditions, the same
type of variational structure can lead to different physical
conclusions.

\section{\normalsize{Summary and outlook}}
%\label{three}
%%%%%%%%%%%%%%%%%%%%%%%%%%%%%%%%%%%%%%%%%%%%%%%%%%%%%%%%%%%%%%%%%%%%%%%

In Sects.2 and 3, I have illustrated an analytic, variational
calculation of the energy spectrum for the broken-symmetry phase of
the basic $\Phi^4$ Hamiltonian (\ref{normal}). In a continuum limit
where the ultraviolet cutoff $\Lambda\to \infty$ and the coupling
constant $\lambda \to 0$, such that $\lambda \ln\Lambda$ is finite,
the variationally determined spectrum $E_1({\bf{p}})$ approaches the
free-field form at large $|{\bf p}|$, namely \BE E_1({\bf{p}}) \to
\sqrt{ {\bf p}^2 + m^2_h} \left(1 + {\cal O}( {{m^2_h}\over{ {\bf
p}^2 }}) \right) \EE However, in the same continuum limit, the
energy-gap $E_1(0)$ remains definitely smaller than the $m_h$
parameter that controls the asymptotic shape of the spectrum. With
the trend for the coupling coupling constant in Eq.(\ref{choice}),
which is self-consistently determined by the overall minimization of
the effective potential, one finds \BE {{E_1(0)}\over{m_h}}=0.7
\left(1 + {\cal O}({{1}\over{\ln \Lambda}})\right) \EE and the
simple leading behaviour Eq.(\ref{leading}). The variational nature
of the result implies that, by enlarging the subspace to include
higher-order $|b^{\dagger}b^{\dagger}b^{\dagger}\rangle$,
$|b^{\dagger}b^{\dagger}b^{\dagger}b^{\dagger}\rangle$,...contributions
in the Fock space, the ratio ${{E_1(0)}\over{m_h}}$ can only
decrease.

A possible objection might concern the simplest form
Eq.(\ref{normal}) adopted for the Hamiltonian operator. Would the
variational result persist by employing for the contact interaction
more sophisticated de-singularized operators as, for instance, the
generalized normal-ordering prescriptions of Ref.\cite{zimmermann} ?
There is no obvious answer to this question. By replacing the
Hamiltonian operator $\hat{H}$ of Eq.(\ref{normal}) with a new
operator, say $\hat{H}'$, one should first repeat the whole
stability analysis within the class of the gaussian ground states
and later check the consistency of $\hat{H}'$ with the variational
calculation in the $|b^{\dagger}\rangle$ and
$|b^{\dagger}b^{\dagger}\rangle$ sectors. In our case, by using
Eq.(\ref{normal}) (or equivalently the `bare mass' in
Eq.(\ref{bare})), one obtains finite results at all stages, once
Eq.(\ref{choice}) is used self-consistently to determine the cutoff
dependence of the coupling constant. For this reason, the operator
$\Delta \hat{H}=\hat{H}'-\hat{H}$ should only introduce non-leading
divergent terms in the calculations, at least if the trend $\lambda
\sim 1/\ln\Lambda$ has to be maintained.

Therefore, one is naturally driven to interpret the peculiar
infrared behaviour of the broken phase as a true physical effect due
to the existence of a non-trivial vacuum condensate associated with
the typical scale $m_h$. When the momentum increases, the
differences with the trivial empty vacuum become unimportant and the
energy spectrum approaches a standard massive form with $m^2_h\sim
\lambda v^2$. However, when ${\bf p} \to 0$, the presence of the
condensate cannot be reabsorbed into the mass term alone due to the
strong attraction among the bare massive states which is induced by
the cubic interaction proportional to $\lambda v$. For this reason,
it becomes important to understand how fast $E_1(0)$ decreases by
improving on the variational procedure and, in particular, whether
it remains non vanishing in the $\Lambda \to \infty$ limit.

This aspect is closely related to the comparison with the lattice
data mentioned in the Introduction and deserves additional comments.
In general, one can express the inverse connected propagator as \BE
G^{-1}(p) = p^2 +M^2(p^2) \EE If the single-particle spectrum
approximates the form $\sqrt{ {\bf p}^2 + m^2_h}$ at large
$|{\bf{p}}|$ and tends to $E_1(0)< m_h$ when ${\bf{p}}\to 0$, one
can imagine various interpolating forms for $M(p^2)$ in Euclidean
space but, in any case, one expects the propagator to deviate from
the simple form $1/(p^2+m^2_h)$ by approaching the $p_\mu \to 0$
limit. These deviations can be parameterized by using Stevenson's
sensitive variable \cite{stevensonnp} \BE \zeta(p,m)\equiv (p^2
+m^2)G(p) \EE In terms of this variable, by introducing the mass
value $m\sim m_h$ that well describes the high-momentum propagator
data, one gets from Ref.\cite{further} a zero-momentum value \BE
\zeta(0,m_h)= m^2_h G(p=0)> 1 \EE The data also indicate that, by
approaching the continuum limit of the lattice theory,
$\zeta(0,m_h)$ becomes larger and larger while the deviations from
$\zeta \sim 1$ are also confined to a smaller and smaller region of
momenta near $p_\mu=0$ (compare Figs. 3, 4 and 5 of
Ref.\cite{further}). Thus, in the continuum limit, both the
"zero-momentum mass" $\sqrt {G^{-1}(p=0)}$ and the peculiar infrared
region $|p| \lesssim \delta$ where the propagator deviates from the
simple massive form, might vanish in units of the higher-momentum
parameter $m_h$. In this scenario there would be a hierarchy of
scales $\delta \ll m_h \ll \Lambda$ such that
${{\delta}\over{m_h}}\to 0$ when ${{m_h}\over{\Lambda}} \to 0$ (as
for instance with the relation $\delta\sim m^2_h/\Lambda$). However,
if ${{G^{-1}(p=0)}\over{m^2_h}} \to 0$, also $E_1(0)/m_h \to 0$ and
thus both the range $|{\bf p}| \lesssim \delta$ and the
corresponding portion of the energy spectrum $E_1({\bf p})$ would
shrink to the zero-measure set $p_\mu=0$. In this picture of the
continuum limit, where $m_h$ can be taken to define the unit mass
scale, the energy spectrum becomes discontinuous, namely $E_1({\bf
p})=\sqrt{{\bf p}^2 + m^2_h}$ for ${\bf p}\neq 0$ and $E_1({\bf
p})=0$ for ${\bf p}= 0$. Notice that this would represent a
Lorentz-invariant decomposition because the value $(E_1=0,{\bf p}=
0)$ or  $p_\mu=0$ forms a Lorentz-invariant subset.

This point of view agrees well with Stevenson's recent analysis
\cite{recent} of the propagator in the broken-symmetry phase. In his
view, a more faithful representation of the continuum limit can be
obtained by starting from the non-local action \BE \int d^4x\int
d^4y ~\Phi^2(x) U(x-y)\Phi^2(y) \EE The kernel $U(x-y)$ contains,
besides the repulsive contact $\delta-$function term, say $U_{\rm
core}(x-y)$, also an effective long-range attraction for $x \neq y$,
say $U_{\rm tail}(x-y)$. The latter, which is essential for a
physical description of spontaneous symmetry breaking as a true
condensation process \cite{mech}, originates from {\it
ultraviolet-finite} parts of higher-order Feynman graphs and has
never been considered in the perturbative RG$-$approach. Instead, by
taking into account both $U_{\rm core}$ and $U_{\rm tail}$ (and
avoiding double counting), one can define a modified RG$-$expansion
\cite{recent}, as in a theory with two coupling constants. In the
end, by taking the $\Lambda \to \infty$ limit, the resulting
connected Euclidean propagator $G(p)$ has the standard massive form
$G^{-1}(p)=(p^2+m^2_h)$ except for a discontinuity at $p_\mu=0$
where $G^{-1}(p=0)=0$. This type of structure, implying the
existence of a branch of the spectrum whose energy $E_1({\bf p})\to
0$ in the ${\bf p}\to 0$ limit, would indeed support the previous
idea that, at least for $\Lambda \to \infty$, the exact result is
$E_1(0)=0$.

Finally, this discontinuous nature of $G^{-1}(p=0)$ would also be in
agreement with the analogous indication of Ref.\cite{consoli} that,
in the broken-symmetry phase and quite independently of the
Goldstone phenomenon, the zero-momentum connected propagator of the
shifted fluctuation field is a two-valued function that, in addition
to the standard value $G^{-1}_a(p=0)=m^2_h$, includes the solution
$G^{-1}_b(p=0)=0$ as in a massless theory. It is conceivable that
such a subtle, nearly point-like, effect around $p_\mu=0$ might have
been missed in most conventional approximation schemes. At the same
time, the idea of an infrared sector which is richer than expected
might have far reaching phenomenological implications. For instance,
by using the general properties of the Fourier transform, any
$G^{-1}(p)$ that smoothly interpolates in an infinitesimal momentum
region $| p| \sim \delta\ll m_h$, between $G^{-1}_b(p=0)=0$ and
$G^{-1}_a(p)\sim (p^2+m^2_h)$, would yield a long-range $1/r$
potential of infinitesimal strength $\delta^2/m^2_h$ \cite{plb09}.

In conclusion, for the conceptual relevance of the problem and its
potential phenomenological implications, it seems worth to sharpen
our understanding of the low-momentum region of spontaneously broken
$\Phi^4$ theories both analytically and with a new generation of
numerical simulations on those very large 4D lattices (e.g. $100^4$)
that should now be available with the present computer technology.

\vskip 20 pt

\vfill\eject

\end{document}